\documentclass[%
 aps,
 jmp,%
 amsmath,amssymb,
 reprint,%
]{revtex4-2}

\usepackage{enumitem}
\usepackage{braket}
\usepackage{float}
\usepackage{graphicx}% Include figure files
\usepackage{dcolumn}% Align table columns on decimal point
\usepackage{bm}% bold math
\usepackage{mathtools}
\usepackage{amsthm,amssymb, hyperref}
\usepackage{xcolor}
\usepackage{todonotes}
\usepackage{tikz}
\usetikzlibrary{quantikz}
\DeclarePairedDelimiter{\ceil}{\lceil}{\rceil}

\begin{document}

\preprint{AIP/123-QED}

\title[A scalable quantum gate-based implementation for quantum causal hypothesis testing]{A scalable quantum gate-based implementation for causal hypothesis testing}% Force line breaks with \\

% scalable practical realistic 

% \title[A quantum circuit implementation for causal hypothesis testing]{A quantum circuit implementation for causal hypothesis testing}% Force line breaks with \\

% \title[Quantum Accelerated Causal Tomography...]{Quantum Accelerated Causal Tomography: \\Circuit Considerations Towards Applications}% Force line breaks with \\
%
\author{Akash Kundu}
\email{akundu@iitis.pl}
\thanks{corresponding author}
% \affiliation{QWorld Association.}
\affiliation{Institute of Theoretical and Applied Informatics, Polish Academy of Sciences,}
\affiliation{Joint Doctoral School, Silesian University of Technology, Gliwice, Poland}
% \affiliation{Quantum Machine Learning research group, Quantum Computing division, QuTech, The Netherlands}
\affiliation{Quantum Intelligence Research Team, Department of Quantum \& Computer Engineering, Delft University of Technology, 2628 CD Delft, The Netherlands}
\author{Tamal Acharya}
\email{tamal\_974@yahoo.com}
% \affiliation{QWorld Association.}
% \affiliation{Independent Researcher, Bengaluru, India.}
% \affiliation{Quantum Machine Learning research group, Quantum Computing division, QuTech, The Netherlands}
\affiliation{Quantum Intelligence Research Team, Department of Quantum \& Computer Engineering, Delft University of Technology, 2628 CD Delft, The Netherlands}
\author{Aritra Sarkar}
\email{a.sarkar-3@tudelft.nl}
% \affiliation{QWorld Association.}
% \affiliation{Quantum Machine Learning group, QuTech, Delft University of Technology, Delft, The Netherlands}%
\affiliation{Quantum Intelligence Research Team, Department of Quantum \& Computer Engineering, Delft University of Technology, 2628 CD Delft, The Netherlands}
% \affiliation{Quantum Machine Learning research group, Quantum Computing division, QuTech, The Netherlands}
\affiliation{Quantum Machine Learning Group, Quantum Computing Division, QuTech, 2628 CJ Delft, The Netherlands}

\date{\today}% It is always \today, today,

\begin{abstract}
In this work, we study quantum computing algorithms for accelerating causal inference. Specifically, we consider the formalism of causal hypothesis testing presented in [\textit{Nat Commun} 10, 1472 (2019)]. We develop a quantum circuit implementation and use it to demonstrate that the error probability introduced in the previous work requires modification. The practical scenario, which follows a theoretical description, is constructed as a scalable quantum gate-based algorithm on IBM Qiskit. We present the circuit construction of the oracle embedding the causal hypothesis and assess the associated gate complexities. Additionally, our experiments on a simulator platform validate the predicted speedup. We discuss applications of this framework for causal inference use cases in bioinformatics and artificial general intelligence.
\end{abstract}

%Use showkeys class option if keyword
\maketitle

\iffalse
\begin{quotation}
The ``lead paragraph'' is encapsulated with the \LaTeX\ 
\verb+quotation+ environment and is formatted as a single paragraph before the first section heading. 
(The \verb+quotation+ environment reverts to its usual meaning after the first sectioning command.) 
Note that numbered references are allowed in the lead paragraph.
%
The lead paragraph will only be found in an article being prepared for the journal \textit{Chaos}.
\end{quotation}
\fi

\section{\label{s1}Introduction}

Despite the huge success of machine learning~(ML) algorithms based on deep neural networks, these systems are inscrutable black-box models.
This hampers users' trust in the system and obfuscates the discovery of algorithmic biases arising from flawed generative processes that are prejudicial to certain inputs (e.g., racial discrimination).
Explainable artificial intelligence~(XAI)~\cite{marcinkevivcs2020interpretability} focuses on human understanding of the decision from the learned solution as white-box models.
These models provide results that are understandable for domain experts, thus providing transparency, interpretability, and explainability.
XAI algorithms provide a basis for justifying decisions, tracking and thereby verifying them, improving the algorithms, and exploring new facts.
There has been relatively slow progress in XAI, despite realizing its importance as we increasingly automate critical systems.
Early advances in XAI were based on symbolic reasoning systems and truth maintenance systems.
To achieve causal reasoning~\cite{moraffah2020causal}, rule-based learning and logic-based inference systems were proposed.
Methods to address inherent opaque modern methods like deep learning-based neural networks and genetic algorithms include layer-wise relevance propagation and local interpretability.
There exist other ML algorithms (e.g. decision trees, Bayesian classifiers, additive models) that generate interpretable models i.e. the model components (e.g., the weight of a feature, a path in a decision tree, or a specific rule) can be directly inspected to understand the predictions.
However, these models are not general and scalable to compete with the adoption and impact of neural networks.
On the other hand, symbolic reasoning systems were abandoned owing to the difficulty in scaling these systems for a large number of parameters.

The capability of \textit{quantum computation allows us to scale symbolic reasoning models by encoding the classical rules as a superposition of quantum states or processes}~\cite{sarkar2021estimating} is a core motivation in quantum explainable AI.
Quantum mechanics provide enhanced ways to identify causal links; for example, certain quantum correlations can be used to infer classical causal relationships~\cite{ried2015quantum,fitzsimons2015quantum}.
This could overcome the apprehension of existing classical approaches being pursued in XAI.

In this article, we will explore how we can distinguish quantum processes by their causal structure.
Specifically, we study the construction proposed in \cite{chiribella2019quantum} towards a quantum circuit implementation on the IBM Qiskit quantum programming language.
In doing so, we uncover (i) the \textit{implementation aspects of the causal oracle}, (ii) the \textit{gate and qubit complexity} of the full algorithm, and (iii) \textit{practical case error probability}, which we introduce in the paper that shows the dependence of the error probability on some distance measure between the set of processes/hypotheses being tested.
While the current technology readiness level of quantum systems prevents us from demonstrating this causal reasoning within a broader application framework, we present the quantum kernel that can be readily embedded within a software pipeline for applications in bioinformatics and artificial general intelligence~(AGI).
In particular, it will be useful in XAI pipelines~\cite{lavin2021simulation,maruyama2021categorical}.

The remaining article is organized as follows. 
In {Section~[\ref{sec:overview-causal-inference}]} we briefly review the specifics of causal reasoning and some of the well-studied techniques. 
A discussion on the basic concepts and quantum advantage of causal hypothesis testing is given in {Section~[\ref{sec:causal-hypothesis-testing}]}. 
In the following {Section~[\ref{sec:prob-formulation}]} we describe the problem formulation; containing the main findings of the article. 
Here we define a model implementation on Qiskit followed by a correction factor introduced in the error probability based on our empirical results; which we call \textit{practical error probability}. 
Finally, in {Section~[\ref{sec:application framework}]} we discuss some potential use cases in bioinformatics and artificial general intelligence.
The corresponding quantum resources of gates and qubits are assessed for realistic cases.
{Section~[\ref{sec:conclusion}]} concludes the article.

\section{\label{sec:overview-causal-inference}Overview of causal inference}
% \textcolor{red}{\textbf{REMOVE OR MAKE IT TINY!!!!!!!!!!!!!!!!!!!!!!}}
% Identifying cause–effect relations is an essential tool in science.
% Simulation Intelligence: Towards a New Generation of Scientific Methods
% % https://arxiv.org/abs/2112.03235
% This is generally accomplished through classical statistical trials where alternative hypotheses are tested against each other.
% \subsection{Overview of causal inference}
Causal inference %refers to an intellectual discipline that 
considers the assumptions, study designs, and estimation strategies that allow researchers to draw conclusions on the cause-effect relationships between data. 
% The term `causal conclusion' used here refers to a conclusion regarding the effect of a causal variable (often referred to as the ‘treatment’ under a broad conception of the word) on some outcome(s) of interest. 
% The dominant perspective on causal inference in statistics has philosophical underpinnings that rely on the consideration of counterfactual states. 
In particular, it considers the outcomes that could manifest given exposure to each of a set of dynamics of a specific causal variable. %treatment conditions. 
Causal effects are defined as comparisons between these potential outcomes. 
% Causal inference consists of a family of statistical methods whose purpose is to answer the question of why something happens. 
Standard approaches in statistics, such as regression analysis, are concerned with quantifying how changes between two variables are associated, with no directional sense. 
In contrast to that, causal inference methods are used to determine whether changes in a variable X cause changes in another variable Y or vice-versa. 
% Therefore, unlike methods that are concerned with associations only, causal inference approaches can answer the question of why Y changes. 
If X is causally related to Y, then Y's change can (at least partially) be explained in terms of X's change.

% \subsection{Problems with causal inference}
\subsection{Challenges of performing causal inference}

% In this section we discuss the problems with causal inferences and some of the limitations of state-of-the-art causal models.

Causal models are based on the idea of potential outcomes. 
% The fundamental problem for causal inference is that, for any individual unit (a physical object at a particular point in time), we can observe only one of the potential outcomes, either Y(1) (under treatment) or Y(0) (under no treatment). 
% As we observe the value of the potential outcome under only one of the possible treatments~\footnote{A treatment is an action that can be applied or withheld from that unit.}, namely, the treatment actually assigned, hence the potential outcome under the other treatment is missing.
The two major challenges is causal inference are:

% The inadequacy of the causal models is due to their failure to include relevant spatial and structural information in a way that does not render the model non-explanatory, unmanageable, or inconsistent with the basic assumptions of causal graph theory. 
% Making valid causal inferences is challenging because it requires high-quality data and adequate statistical methods. 
% One prominent example of common non-causal methodology is the erroneous assumption of correlative properties as causal properties. 
% Correlated phenomena may not carry inherent causation. 
% Regression models are designed to measure variance within data relative to a theoretical model. 
% This suggests that there is nothing in the data that presents high levels of covariance that have any meaningful relationship among them. 
% This suggests the absence of a proposed causal mechanism with predictive properties or a random assignment of treatment. 
% The presupposition that two correlated phenomena are inherently related is a logical fallacy, known as spurious correlation.
\begin{itemize} %[noitemsep,nolistsep]
\item \textbf{Causation does not imply association}
	For example, we want to compare the impact of an academic degree on the income of a middle-aged individual. 
	The person might have attended the academic degree or might not have. 
	To calculate the causal effect of having an academic degree, we need to compare the output in both situations, which is not possible. 
	This dilemma is a fundamental problem of causal inference. 
	The challenge in causal inference is that \textit{all potential outcomes are not observed}, we only observe one.
	Another example is described by the situation of -- a black cat ran under the fence and I tripped and fell over. 
	We could have tripped anyway. 
	It had nothing to do with the cause of the cat running under the fence, i.e., \textit{associated events do not imply causal connection}.
	
    \item \textbf{Correlation does not prove causality}
	
	Causal inference is the process of drawing a conclusion about a causal connection based on the occurrence of an effect.
        Causal inference is usually a missing data problem~\cite{guo2020survey} and we tend to make \textit{assumptions to make up for the missing causes/variables}. 
	An example is a correlation between people eating ice cream and people drownings. 
	It could indicate that eating ice cream affects drowning. 
	The actual correlation is between the season (summer) and these otherwise unrelated things. 
	In this case, the missing cause is the season. 
	% Another example is the correlation between higher SAT scores and a greater number of books in the house of the student taking the tests. 
	% The causal inference would imply that the number of books directly affects the SAT scores when in reality, they are both affected by something else (in this case most likely a higher average intelligence in the household). 
	
\end{itemize}
% The main difference between causal inference and inference of association is that the former analyzes the response of the effect variable when the cause is changed, and in the latter, inferences about the strength of association between variables are made using a random multivariate sample of data drawn from the population of interest.
% Thus, in causal inference, we have access to the additional information about the variable/s on which the intervention has occurred between two sets of associated data samples.
% Another problem of causal inference is that the cause and the effect may have occurred by chance rather than by intention.

% The causal effect of a drug on systolic blood pressure, one month after the drug dosage vs no exposure to the drug dosage can be represented as a comparison of systolic blood pressure that would be measured at the time of given exposure to the drug with the systolic blood pressure that would be measured at the same point in time in the absence of exposure to the drug.
% The challenge for causal inference is that we are not generally able to observe both of these states: at some point in time when we are measuring the outcomes, each individual either has had drug exposure or has not.
% From the data perspective, there are some open problems to review regarding the great potential of learning causality with data. 
% Such as the study of heterogeneous groups, learning causality with imbalanced data, and learning causality with complex variables.
\subsection{Classical techniques in causal inference}
Causal inference is conducted via the study of systems where the measure of one variable is suspected to affect the measure of another. 
Causal inference is conducted with regard to the scientific method. 
The first step of causal inference is to formulate a falsifiable null hypothesis, which is subsequently tested with statistical methods.
Frequentist statistical inference uses statistical methods to determine the probability that the data occur under the null hypothesis by chance.
Bayesian inference is used to determine the effect of an independent variable.

Common frameworks for causal inference include the causal pie model (component-cause)~\cite{rothman2005causation}, Pearl's structural causal model (causal diagram and do-calculus)~\cite{pearl2000models}, structural equation modeling, and Rubin's causal model (potential-outcome)~\cite{imbens2015causal}. 
\cite{yao2021survey} presents a more comprehensive survey of classical causal inference techniques.

The most frequently used causal models can be grouped into two kinds: causal Bayesian networks and structural equation models (which are distinct but closely related). %, cf. Spirtes 2010
Causal graph models combine mathematics and philosophy: the mathematical elements are Directed Acyclic Graphs (DAGs) and probability theory (with a focus on conditional independence); the philosophical elements are assumptions about the relationship between causation and probability~\cite{spirtes2000causation}.
% \subsection{Algorithmic machine learning}
An alternative approach to causal inference based on \textit{algorithmic generative models} is currently gaining popularity. 
\cite{zenil2019causal} describes the process of performing causal deconvolution using this technique.
This paper talks about the different generating mechanisms by which complex data is produced. 
The authors introduced a universal, unsupervised, and parameter-free model-oriented approach based upon algorithmic probability that decomposes an observation into its most likely algorithmic generative sources. 
This is closely related to the quantum approaches discussed in the next section.

\begin{figure*}[t!]
	\centering %LBRT
	\includegraphics[clip, trim=3cm 0cm 0cm 9cm,width=\textwidth]{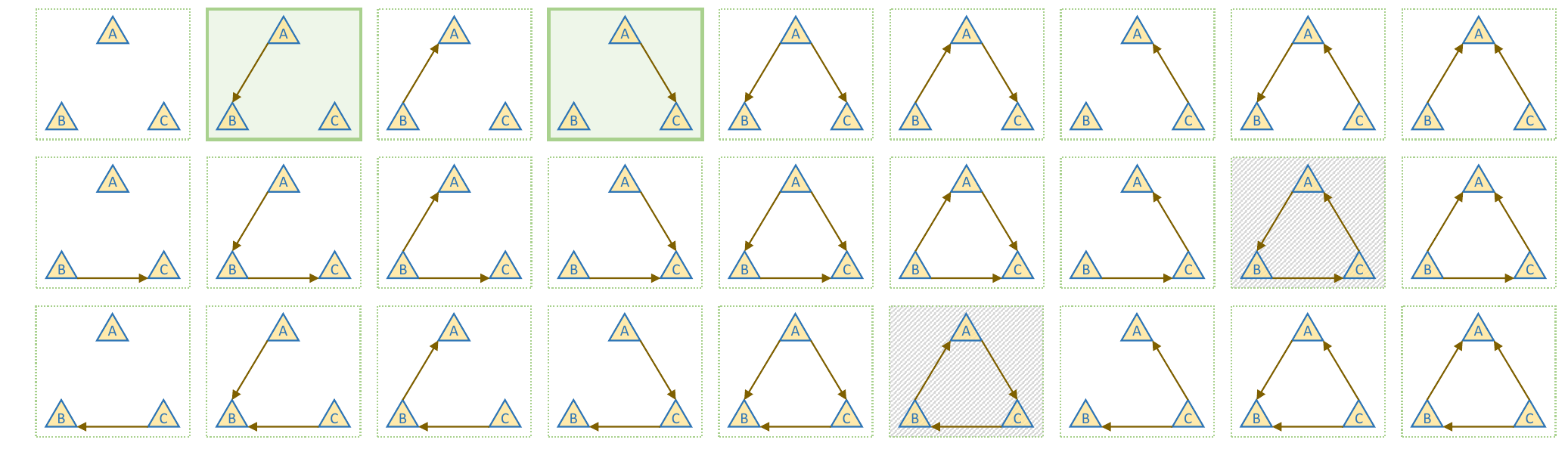}
	\caption{All possible causal relations for 3 variables. Blocks shaded grey have causal loops (and are typically not considered). The set of blocks shaded green indicates a case of causal hypothesis testing.}
	\label{fig:3vci}
\end{figure*}

\subsection{Quantum computation and algorithmic generative models}

The synergy between quantum computation and algorithmic information has been studied extensively in \cite{sarkar2022applications}.
Two main directions were explored, that can be applied for causal inference.

In \cite{sarkar2021estimating} a global/objective view is presented, which involves quantum automata for algorithmic information.
A framework for causal inference based on algorithmic generative models is developed. 
This technique of quantum-accelerated experimental algorithmic information theory~(QEAIT) can be ubiquitously applied to diverse domains. 
Specifically for genome analysis, the problem of identifying bit strings capable of self-replication is presented. 
A new quantum circuit design of a quantum parallel universal linear bounded automata~(QPULBA) model is developed that enables a superposition of classical models/programs to be executed, and their properties can be explored. 
The automaton prepares the universal distribution as a quantum superposition state which can be queried to estimate algorithmic properties of the causal model.

In \cite{sarkar2021qksa} a local/subjective view is presented, which involves universal reinforcement learning in quantum environments.
This theoretical framework can be applied to automated scientific modeling. 
A universal artificial general intelligence formalism is presented that can model quantum processes. 
The developed quantum knowledge-seeking agent~(QKSA) is an evolutionary general reinforcement learning model for recursive self-improvement. 
It uses resource-bounded algorithmic complexity of quantum process tomography~(QPT) algorithms. 
The cost function determining the optimal strategy is implemented as a mutating gene within a quine. 
The utility function for an individual agent is based on a selected quantum distance measure between the predicted and perceived environment.

These techniques motivate our research in this article.
Unlike quantum/classical data-driven machine learning, these primitives of QEAIT and QKSA preserve the explanatory power of the model the learning converges to by exploring the space of programs on an automata model.
The QPULBA model (in QEAIT) and QPT algorithms (in QKSA) can be generalized to a causal oracle and causal tomography respectively.

Causal approaches for quantum machine learning (QML)~\cite{steinmuller2022explainable,heese2023explainable}, and quantum algorithms for causal inference~\cite{burge2023quantum,pira2023explicability} are also a related active research direction.

\section{Causal hypothesis testing} \label{sec:causal-hypothesis-testing}

A canonical approach in causal inference is to formulate different hypotheses on the cause–effect relations and test them against each other.
This technique is typically used when there is some knowledge of the characteristic of the phenomenon that is being tested.
The complete search space of directed graphs between events grows exponentially.
An exhaustive example of 3 variable causal relations is shown in Fig.~[\ref{fig:3vci}].
In causal hypothesis testing~(CHT) a subset of these graphs is considered as the set of hypotheses being tested against each other.
For example, in Fig.~[\ref{fig:3vci}], we can consider, two hypotheses (shaded in green):
\begin{itemize}[nolistsep,noitemsep]
	\item A causes B, C is independent
	\item A causes C, B is independent.
\end{itemize}

\subsection{Quantum advantage in classical causal hypothesis testing}

Quantum information enables a richer spectrum of causal relations that is not possible to access via classical statistics.
Most research in this direction is towards exploring causality in the quantum context~\cite{costa2016quantum,giarmatzi2019quantum,javidian2021quantum,bai2021quantum,bai2020efficient,bavaresco2021strict}.
Our focus in this work is specifically using the quantum formulation to provide a \textit{computational advantage with respect to a classical technique on classical data}.

This problem is studied extensively in \cite{chiribella2019quantum}.
The research analyzes the task of identifying the effect of a given input variable. 
In the full quantum version of the problem the variables A, B and C are considered as a quantum system of dimension $d$ which in return satisfies either of the two following causal hypotheses
(1) $B$ resulted from $A$ through an arbitrary unitary operation and the state of $C$ is maximally mixed, or (2) $C$ resulted from $A$ through an arbitrary unitary operation and $B$ is maximally mixed. The set of allowed causal relationships say $\mathcal{C}$, between input and output, depends on the physical theory, which determines which maps can be implemented by physical processes. In classical physics, cause-effect relations are typically represented by conditional probability distributions, while, in quantum theory, they are described by quantum channels, i.e., completely positive trace-preserving maps transforming density matrices of an input quantum system $A$ into density matrices of an output (either $B$ or $C$).

Despite the fact that a cause-effect relationship can be established utilizing any unitary operation, the error probability that is attained remains
\begin{equation}
    p_{err}=\frac{1}{2d^N},
    \label{eq:error-prob-quantum-advantage1}
\end{equation}
where $N$ represents the number of interventions between one instance and the next.
\begin{figure}[bht]
    \centering
    \includegraphics[width=\columnwidth]{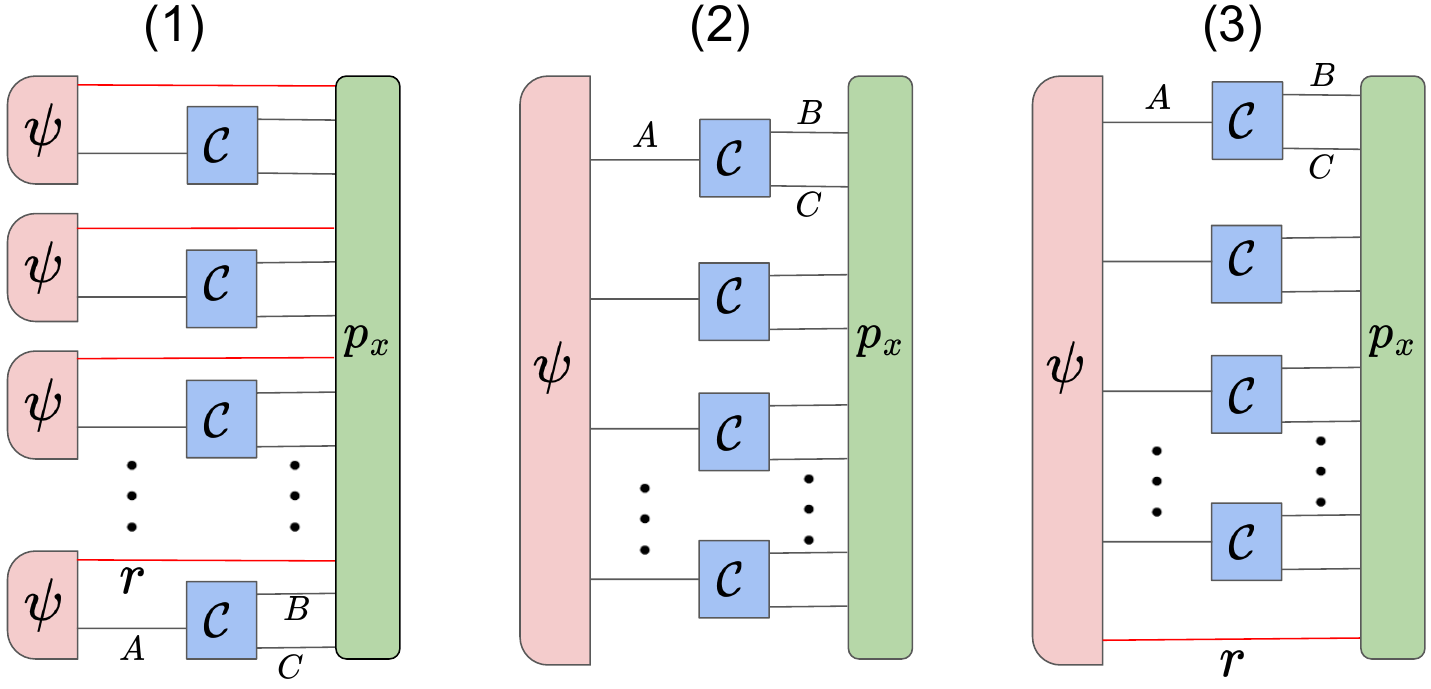}
    \caption{Here we illustrate the two fundamental ways of constructing parallel strategy. In (1) we initialize the quantum system $A$ along with a reference $r$ in a state $\psi$. This setting is repeated for $N$ times. Meanwhile in (2) an $N$ size input, $\psi$, is provided as an input $A$. Similar to the previous strategy in (3) we have the same $N$ probes of the input entangled using an additional reference $r$. 
    An unknown process $\mathcal{C}$ (an arbitrary unitary) induces a causal relation between the input and the output in all settings.}
    \label{fig:parallel_strategy}
\end{figure}

The error probability in Eq.\ref{eq:error-prob-quantum-advantage1}
is $d$ times smaller than the classical error probability. To achieve this advantage the authors in \cite{chiribella2019quantum} make use of a universal quantum strategy by preparing a $d$ particle singlet state of the form
\begin{equation}
    \ket{s_d} = \frac{1}{\sqrt{d!}}\sum_{i_1,i_2,\ldots,i_d}\epsilon_{i_1,i_2,\ldots,i_d}\ket{i_1},\ket{i_2},\ldots,\ket{i_d}\label{eq:universal-quantum-strategy},
\end{equation}
where $\epsilon_{i_1,i_2,\ldots,i_d}$ is an asymmetric tensor and the sum
ranges over all vectors on the computational basis. Then, each of
the $d$ subsystems are fed as an input to one use of the channel $\mathcal{C}$. By repeating the experiment for $t$ times, and by performing Helstrom’s
minimum error measurement one can attain the error probability given in Eq.~\ref{eq:error-prob-quantum-advantage1} with $N=t\times d$. This exact strategy is illustrated in Fig.~\ref{fig:parallel_strategy}(1).

\begin{figure}[htb]
	\centering
	\includegraphics[width = 0.99\columnwidth]{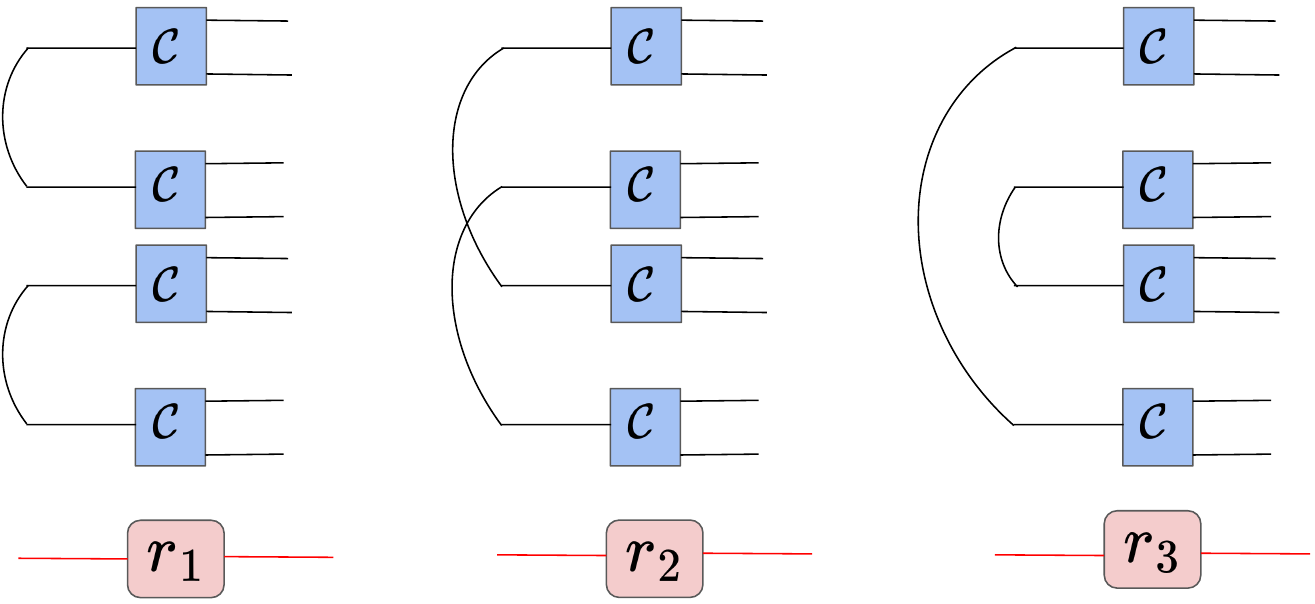}
	\caption{Here we illustrate the different ways of partitioning a quantum system of $N=4$ size into $d=2$ groups. The three distinct configurations are superposed in the input state $\psi$ and can be accessed by the state of the reference $r$.}
	\label{fig:hypo_probe}
\end{figure}

Apart from the error probability, the discrimination rate, $R$, is a very crucial performance quantifier for causal hypothesis testing protocols. It can be defined as the rate at which the two causal hypotheses
can be distinguished from each other. In Fig.~\ref{fig:parallel_strategy} for the strategy (1) and (2) the discrimination rate remains $\log d$. But slight engineering of the protocol (2) that adds a reference $r$ (see protocol (3) of Fig.~\ref{fig:parallel_strategy}) helps us to achieve a discrimination rate of $2\log d$, which is twice as fast as the optimal classical strategy. 
The primary motive behind the introduction of the reference $r$ lies in the fact that it helps entangle the $N$ input probe states. 
To say in a more elaborate manner, in the absence of the reference we saw through strategy (2) of Fig~\ref{fig:parallel_strategy} that it is optimal to partition the $N$ input into $\frac{N}{d}$ groups and then entangle the probe with each group. 
Generalizing this line of thought, we consider a mechanism where the partition of the subsystems happens according to a certain configuration $i$ if a control system is in the
state $\ket{i}$. 
Thus, when the control system is in a superposition, the optimal input state is
\begin{equation}
    \ket{\psi} = \frac{1}{n}\sum_{i=1}^{n}\left(\ket{s_d}^{\otimes \frac{N}{d}}\right)_i\otimes\ket{i},
    \label{eq:parallel_correlated_input}
\end{equation}
where $i$ is labelled for different ways to partition $N$ indistinguishable objects into groups of $d$ elements, the $n$ is the number of such ways of partitioning, and
$\ket{s_d}^{\otimes \frac{N}{d}}$ are orthogonal states of the reference system represented through a product of $\frac{N}{d}$ singlet states. $s_d$ is defined through Eq.~\ref{eq:universal-quantum-strategy}. In Fig.~\ref{fig:hypo_probe} the different ways of dividing $N=4$ copies of the causal process probes ($\mathcal{C}$), which yields $3$ different ways of entangling bi-partitions of these probes, indexed by encoding on the reference register (through $r_i$, $i=1,2,3$).

The above observation helps us to conclude that the quantum correlation speeds up the causal hypothesis testing. However, the caveat of the approach is that it is not practically feasible due to the resource scaling of this encoding. The choice of a maximally mixed quantum channel as an alternate hypothesis prevents the step toward the practicality of the approach. In this work, we tackle this issue by easing the previously proposed general CPTP scenario in favor of \textit{constructable causal oracles}.
We provide a scalable quantum algorithm that can be implemented in a gate-based quantum computer. This approach is described briefly in the next section.
% }
% {The error probability of the causal hypothesis testing is defined as the fraction of the misidentified hypothesis.}
% The constructed quantum strategy reduces the error probability by an exponential amount, doubling the decay rate of the error probability with the number of accesses to the relevant variables. 
% This decay rate is the highest achievable rate allowed by quantum mechanics, even if one allows for exotic setups where the order of operations is indefinite.
% The key ingredient of the quantum speedup is the ability to run multiple equivalent experiments in a quantum superposition.
% This can be used for the detection of a causal link between two variables, and in the identification of the cause of a given variable, as well.

% The set of allowed causal relationships depends on the physical theory, which determines which maps can be implemented by physical processes.
% In classical physics, cause-effect relations are typically represented by conditional probability distributions, while, in quantum theory, they are described by quantum channels, i.e., entirely positive trace-preserving maps transforming density matrices of system A into density matrices of system B.

\section{Practical implementation considerations}
\subsection{The problem statement} \label{sec:prob-formulation}
Contrary to the general case considered in~\cite{chiribella2019quantum}, which is briefly described in the previous section, our formulation is towards an implementable quantum circuit that is demonstrated on the \texttt{Qiskit} quantum simulator. This formulation is detailed in the following.

As we have already discussed the parallel strategy with quantum correlated inputs through a reference, as shown in Fig.~\ref{fig:parallel_strategy}(3), gives us a quantum advantage in the detection of cause and effect. Hence, we provide step-wise modification to this optimal structure to finally implement it in a quantum computer. In our implementation, we consider the input variables (causes) to be denoted by the set $C = \{c_0, c_1,\dots, c_{|C|-1}\}$, while the output variables (effects) are denoted by the set $E = \{e_0, e_1,\dots, e_{|E|-1}\}$.
Furthermore, we consider an equal number of input and output variables to implement a to preserve unitarity, i.e., $|C| = |E| = k$. 
We maintain the simplification in \cite{chiribella2019quantum}, that all variables are of equal length, 
\begin{equation}
d = |c_i| = |e_j|,
\end{equation}
where $c_i \in C$ and $e_j \in E$.
Thus, each variable has $2^d$ states. 
Furthermore, we consider that each effect is a permutation of only one cause, i.e., the map from $C$ to $E$ is a bijective function, one-to-one correspondence, or invertible function. To make our implementation more general, we consider a scenario where the access to interventions (i.e., $\mathcal{C}$) on the causes and the effects are not equal.

As a proof-of-concept, we implement a causal hypothesis testing with control over $1$ cause $C=\{c_0\}$ and measurement capability over $2$ potential effects $E = \{e_0,e_1\}$.
These hypotheses are mutually exclusive, i.e., 
\begin{itemize}
    \item Hypothesis 0: $e_0 \leftarrow U c_0$, $e_1$ is independent of $c_0$,
    \item Hypothesis 1: $e_1 \leftarrow U c_0$, $e_0$ is independent of $c_0$,
\end{itemize}
where $U$ is a unitary map between the cause and effect. Describing all the necessary ingredients for causal hypothesis testing, we now head towards practical implementation.

\subsection{The implementation}
Here we take the parallel strategy as described in Fig.~\ref{fig:parallel_strategy}(3) and discuss the real-world implementation of its subroutines.
To design a quantum circuit, all variables that are potentially inspected as causes need to be assigned a quantum information placeholder.
There are a total of
\begin{equation}
k = \max\{|C|,|E|\},
\end{equation}
placeholders are required for encoding cause of size $|C|$ and effect of size $|E|$.
Throughout the paper, we consider, $k=2$ sets of qubit registers that are used, namely system $A$ and system $B$.
%-------------------------------------------------

\subsubsection{\textbf{Preparation of $\ket{\psi}$}}

% \textcolor{red}{\paragraph{\textbf{Preparation of $\ket{\psi}$}} 
The very first component of the causal hypothesis testing strategy is the initial state, $\ket{\psi}$, along with the reference $r$. We saw in Eq.~\ref{eq:parallel_correlated_input} that the state is in a superposition of different partitions of the input and the partition depends on $d$. Each state in the superposed control input can be accessed by the state of a control system, which is called the reference $r$. As each partition forms a singlet state of the form Eq.~\ref{eq:universal-quantum-strategy}, we make use of the bell unitary consisting of a Hadamard (\texttt{H}) and \texttt{CNOT} gate to represent the singlet.

\[\centering\begin{quantikz}
&\gate[2]{U_\text{bell}} &\qw \\
& \qw & \qw
\end{quantikz} =
\begin{quantikz}
&\gate{H}&\ctrl{1}&\qw\\
&\qw&\targ{}&\qw
\end{quantikz}\]

To describe in a simpler manner, the wires in Fig.~\ref{eq:parallel_correlated_input} that correlating with different inputs is 
represented by the the $U_\text{bell}$.
\begin{figure}[bth]
    \centering
    \includegraphics{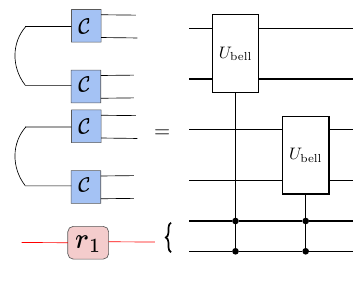}
    \caption{Here we illustrate the practical way to implement one of the partitions of the correlated input with the help of the reference $r$.}
    \label{fig:partition-implementation}
\end{figure}
We illustrate one of the partitions in Fig.~\ref{fig:partition-implementation} where it can be seen that we consider two qubits to represent the reference (i.e. the control system), this is because, for $N=4$ and $d=2$, there are three possible partitions and to trigger each configuration we need at most 3 linear combinations of the control in the form 00, 01, 10. This can be achieved using at least two qubits initialized with $\texttt{H}\otimes\texttt{H}$. The full circuit for all the combinations is illustrated through $U_\text{per}$ in Fig.~\ref{fig:full-perm-circuit}.

For $n$-qubits, the total possible such linearly independent pair combinations are $r = \dfrac{n!}{(n/2)! \times 2^{n/2}}$, where
\begin{equation}
    n = N\times|C|\times d.
\end{equation}
Since we have control over only $1$ cause the $n = 4\times1\times1 = 4$, and $r = 4!/(2!\times2^2) = 3$. Meanwhile, the number of qubits in the reference to encode these states is $N_{\textrm{ref}} = \lceil\log_2 r \rceil$ and in this case $N_{\textrm{ref}} = \lceil \log_2 3 \rceil = 2$.

\subsubsection{\textbf{Preparation of $\mathcal{C}$}}

\begin{figure}[H]
    \centering
\begin{quantikz}
 \lstick[4]{System A}&\swap{4} &\qw & \qw & \qw & \qw\\
 & \qw & \swap{4} & \qw & \qw & \qw\\
 & \qw & \qw & \swap{4} & \qw & \qw\\
 & \qw & \qw & \qw & \swap{4} & \qw\\
 \lstick[4]{System B} & \targX{} & \qw & \qw & \qw & \qw\\
 & \qw & \targX{} & \qw & \qw & \qw\\
 & \qw & \qw & \targX{} & \qw & \qw\\
 & \qw & \qw & \qw & \targX{} & \qw
\end{quantikz}
    \caption{Illustration of Eq.~\ref{eq:oracles_1} for system $A$ (the first 4 qubits) and $B$ (the last 4 qubits) of size $N=4$.}
    \label{fig:ill-orc1}
\end{figure}
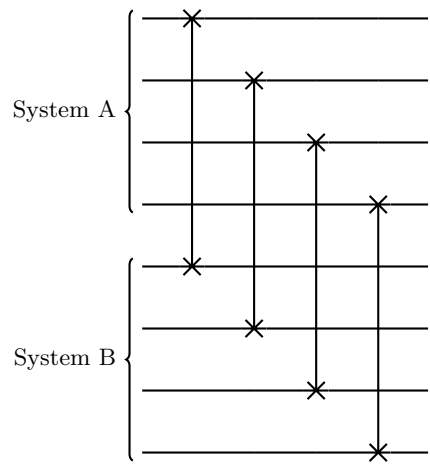

Recalling that $\mathcal{C}$ induces a causal relationship between the input and output and in quantum, it is defined as a quantum channel that maps the input density matrices to an output. To be consistent with the \textit{Hypothesis 0} and \textit{Hypothesis 1}, described in section~\ref{sec:prob-formulation} we choose $U = \mathbb{I}$ for the \textit{Hypothesis 0} and a $U=\texttt{SWAP}$ operation for \textit{Hypothesis 0}. The unitary that preserves the essence of the \textit{Hypothesis 1} is given by
\begin{equation}
U_\textrm{orc}^{H_{1}} =
\textrm{SWAP}(A_i,B_i)^{\otimes N}\otimes\mathbb{I}(A_i)^{\otimes N}\otimes \mathbb{I}(B_i)^{\otimes N},
\label{eq:oracles_1}
\end{equation}
and shown in Fig.~\ref{fig:ill-orc1}.

\begin{figure}[H]
    \centering
\begin{quantikz}
\lstick[4]{System A} & \qw &\swap{4} &\qw & \qw & \qw & \qw\\
& \qw & \qw & \swap{4} & \qw & \qw & \qw\\
& \qw & \qw & \qw & \swap{4} & \qw & \qw\\
& \qw & \qw & \qw & \qw & \swap{4} & \qw\\
\lstick[4]{System B} & \qw & \targX{} & \qw & \qw & \qw & \qw\\
& \qw & \qw & \targX{} & \qw & \qw & \qw\\
& \qw & \qw & \qw & \targX{} & \qw & \qw\\
& \qw & \qw & \qw & \qw & \targX{} & \qw\\
\lstick{Ancilla} & \gate{RX(\theta_\text{ctrl})} & \ctrl{-5} & \ctrl{-5} & \ctrl{-5} & \ctrl{-5} & \qw
\end{quantikz}
    \caption{Here we illustrate the Hypothesis presented in Eq.~\ref{eq:oracles_2} for system $A$ (the first 4 qubits) and $B$ (the last 4 qubits) of size $N=4$.}
    \label{fig:ill-orc2}
\end{figure}
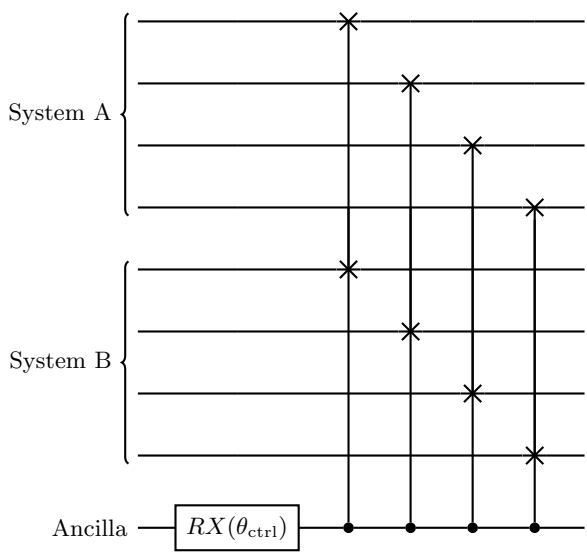

Now to merge the $\mathbb{I}$ and the \texttt{SWAP} \textit{Hypotheses} in s single unitary and to gain better control over the applied hypothesis, we modify Eq.~\ref{eq:oracles_1} by introducing an ancillary qubit initialized with \texttt{RX($\theta_\text{ctrl}$)}. The ancilla qubit works as the control of the SWAP that for $\theta_\text{ctrl}=0$ behaves as $\mathbb{I}$ and for $\theta_\text{ctrl}=\pi$ gives complete $\texttt{SWAP}$ operation. The illustration of the hypotheses is provided in Fig.~\ref{fig:ill-orc2} and given by Eq.~\ref{eq:oracles_2}. 
\begin{widetext}
\begin{equation}
U_\textrm{orc}^{H_{0/1}}(\theta_\text{ctrl}) = \big[\textrm{CSWAP}(q_\textrm{anc},A_i,B_i)^{\otimes N}\big]
\big[\textrm{RX}(q_\textrm{anc},\theta_\textrm{ctrl})\otimes\mathbb{I}(A_i)^{\otimes N}\otimes \mathbb{I}(B_i)^{\otimes N}\big],
\label{eq:oracles_2}
\end{equation}
\end{widetext}

\subsubsection{\textbf{Initial choice of system A, B}}

We have already mentioned that we use $k=2$ sets of qubit registers to encode the cause and effects through systems A and B. 
As our work primarily focuses on how different hypotheses affect the error probability $p_\text{err}$, we restrict ourselves to the all-zero, i.e., $\ket{0}^{\otimes N}$ initialization. 
But it should be noted that the causes and effects need not be the vacuum states but can be any arbitrary quantum state. 
To introduce this degree of freedom, we propose two unitaries $U_\text{in}^A$ and $U_\text{in}^B$ that randomly initialize the systems A and B.

\begin{figure*}[bht]
	\centering
	\includegraphics[width = \linewidth]{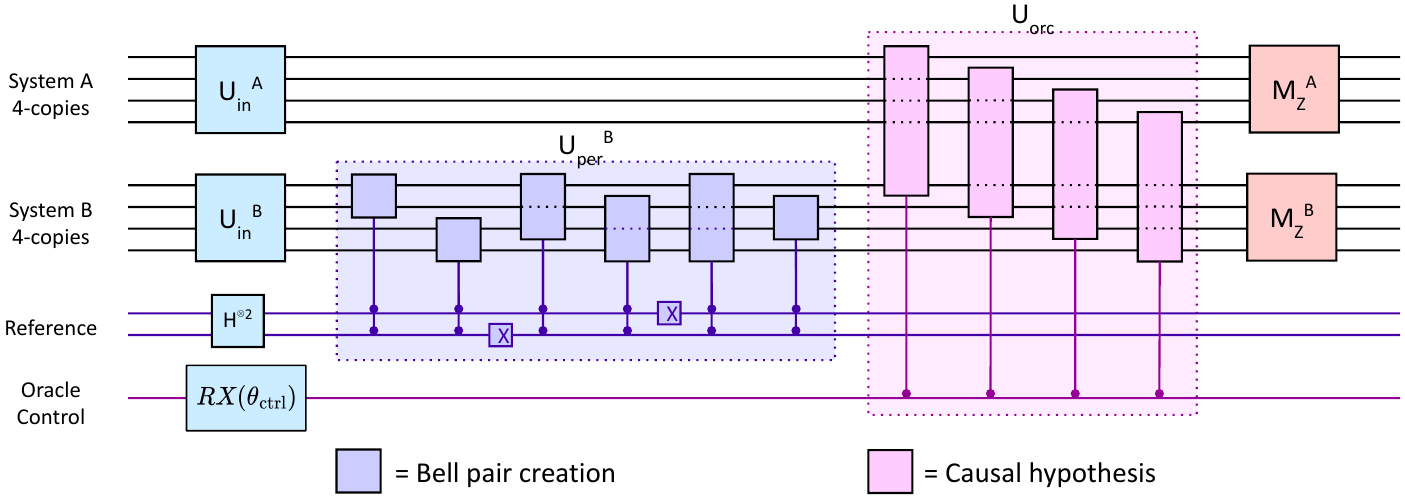}
	\caption{The illustration of the model implementation. Where we also show the decomposition of $U_\textrm{per}$ for $ N_\textrm{A} = N_\textrm{B} = 4$.}
	\label{fig:full-perm-circuit}
\end{figure*}

\subsubsection{\textbf{Measuring the outcome}}

In the original work, Helstorm's minimum error measurement~\cite{helstrom1969quantum} is used to obtain the advantage over the classical causal inference as shown in Eq.~\ref{eq:error-prob-quantum-advantage1}. To calculate the distinguishing probability in computational basis, we add the probabilities of basis states that are unique to one of the qubits of effects (systems A and B).

%----------------

% \begin{figure*}
%     \centering
% \begin{quantikz}
% \lstick[4]{System A} & \gate[4]{U_\text{per}^\text{A}} & \qw & \qw & \qw & \qw & \qw & \qw &\swap{4} &\qw & \qw & \qw & \qw\\
% & \qw & \qw &  \qw & \qw & \qw & \qw & \qw & \qw & \swap{4} & \qw & \qw & \qw\\
% & \qw & \qw & \qw & \qw & \qw & \qw & \qw & \qw &  \qw & \swap{4} & \qw & \qw\\
% & \qw & \qw & \qw & \qw & \qw & \qw & \qw & \qw & \qw & \qw & \swap{4} & \qw\\
% \lstick[4]{System B} & \gate[4]{U_\text{per}^\text{B}} & \gate[2]{U_\text{bell}} & \qw & \qw & \qw & \qw & \qw & \targX{} & \qw & \qw & \qw & \qw\\
% & \qw & \qw & \qw & \gate[2]{U_\text{bell}} & \qw & \qw & \qw & \qw & \targX{} & \qw & \qw & \qw\\
% & \qw & \qw & \qw & \qw & \qw & \qw & \qw & \qw & \qw & \targX{} & \qw & \qw\\
% & \qw & \qw & \qw & \qw & \qw & \qw & \qw & \qw & \qw & \qw & \targX{} & \qw\\
% \lstick{Reference} & \gate[2]{H^{\otimes 2}} & \qw & \qw & \qw & \qw & \qw & \qw & \qw & \qw & \qw & \qw & \qw\\
% & \qw & \qw & \qw & \qw & \qw & \qw & \qw & \qw & \qw & \qw & \qw & \qw\\
% \lstick{Ancilla} & \gate{RX(\theta_\text{ctrl})} & \qw & \qw & \qw & \qw & \qw & \qw & \ctrl{-7} & \ctrl{-7} & \ctrl{-7} & \ctrl{-7} & \qw
% \end{quantikz}
%     \caption{Here we illustrate the Hypothesis presented in Eq.~\ref{eq:oracles_2} for system $A$ (the first 4 qubits) and $B$ (the last 4 qubits) of size $N=4$.}
%     \label{fig:ill-orc2}
% \end{figure*}
%----------------

The overall quantum circuit to perform the causal hypothesis testing is provided in Fig.~\ref{fig:full-perm-circuit} where (1) $U_\text{in}^A$ and $U_\text{in}^B$ defines the initial choice of system A, B, (2) $U_\text{per}^\text{B}$ prepares $\ket{\psi}$ (see Eq.~\ref{eq:parallel_correlated_input}), (3) The $U_\text{orc}$ presents the hypothesis in Eq.~\ref{eq:oracles_2} and finally (3) $M_Z^A$ and $M_Z^B$ are the computational base measurement to get the error probability $p_\text{err}$.

\begin{figure*}[bht]
	\centering
	\includegraphics[width = 0.7\linewidth]{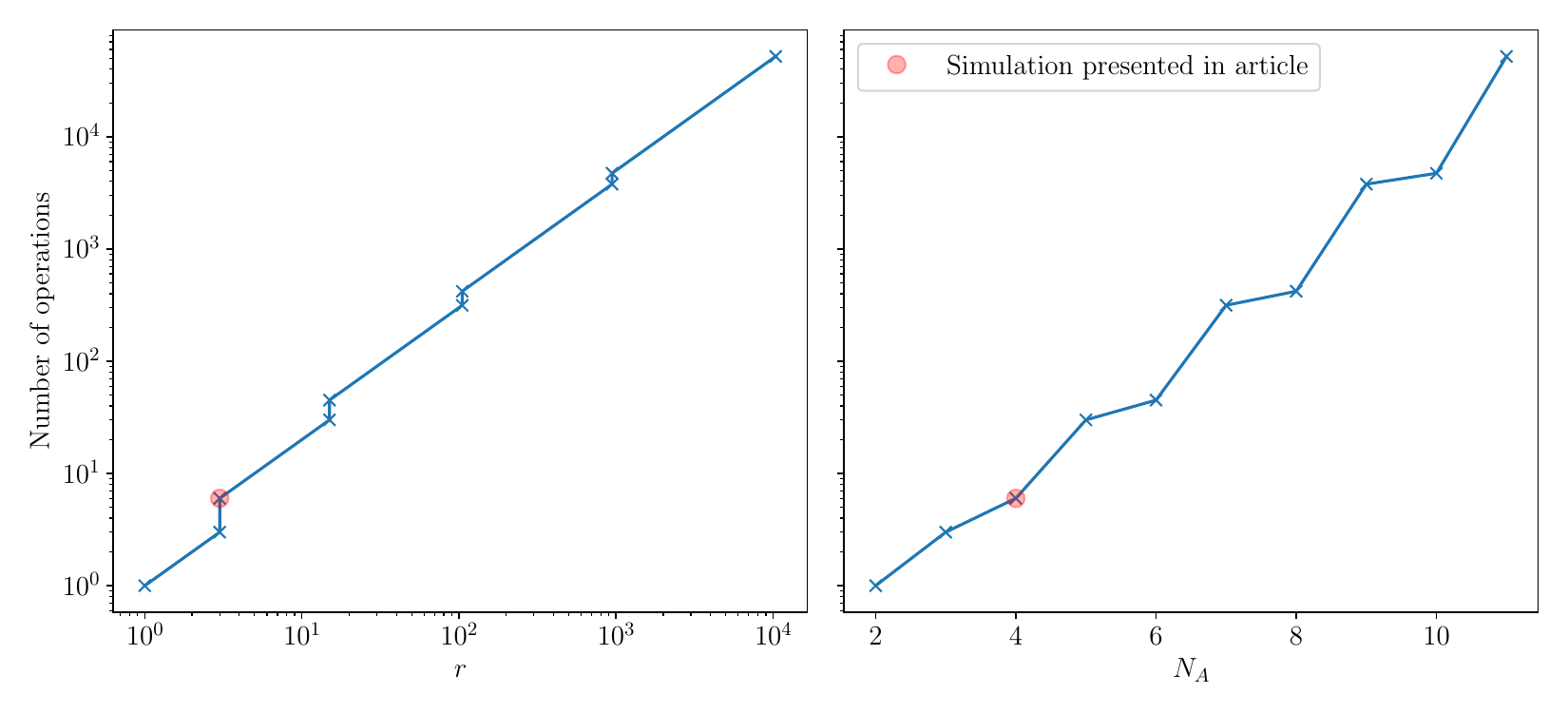}
	\caption{Illustration of the number of controlled Bell unitary required to implement the $U_\textrm{per}$ in respect to a number of linearly independent state $r$ (left-hand side figure) and with the dimension of the subsystem (right-hand side figure).}
	\label{fig:permutation-resource}
\end{figure*}

\subsection{Resource estimation}
The qubit requirement of this model grows as:
\begin{equation}
N \times k \times d + \ceil{\log_2 \left( \dfrac{n!}{\frac{n}{2}! \times 2^{\frac{n}{2}}} \right)}
\label{eq:qubit-requirement}
\end{equation}
The model is illustrated through an implementable quantum circuit in Fig.~[\ref{fig:full-perm-circuit}].
An elaborated discussion of the different components of the quantum circuit can be found in the next section.

Recalling that in the previous work, one of the innovations introduced was to entangle the inputs for the parallel strategy instead of initializing with a tensor product state, which reduces the exponential measurement resource requirement by correlating the basis. However, the cost of implementing the entangled initialization was not analyzed.

Through our model implementation, we are able to show that the operations required to implement the $U_\textrm{per}^B$ grow faster than exponentially with the dimension of the subsystem ($N_A$ or $N_B$) as illustrated in Fig.~[\ref{fig:permutation-resource}] but linear in respect to the linearly independent states, i.e., $r$.

\subsection{Practical error probability}

While the number of controllable causes $|C|$ and the encoding length of the causes $d$ depends on the problem formulation, the number of queries, i.e., $N$ and thereby $r$, is a free parameter.
It can be chosen based on the available quantum circuit resources of the number of qubits in the quantum processor, the decoherence time, and gate error probability, such that the pragmatic error remains low. 
It is shown in ref.~\cite{chiribella2019quantum} that using the correlated input scheme presented in Fig.~\ref{fig:parallel_strategy}(3) we can reach an error probability
\begin{equation} 
p_\textrm{err} = \frac{r}{2d^N}\left( 1-\sqrt{1-r^{-2}} \right) \xrightarrow[]{r>>1}\frac{1}{4rd^N},\label{eq:limiting-case-equation}
\end{equation}
where $r$ is the number of linearly independent states.
It is suggested that the Eq.\eqref{eq:limiting-case-equation} gives the limiting case error probability because, in a more general case, the error probability of causal hypothesis testing should be dependent on the two specific hypotheses.

\begin{figure}[ht!]
	\centering
	\includegraphics[width = 0.8\linewidth]{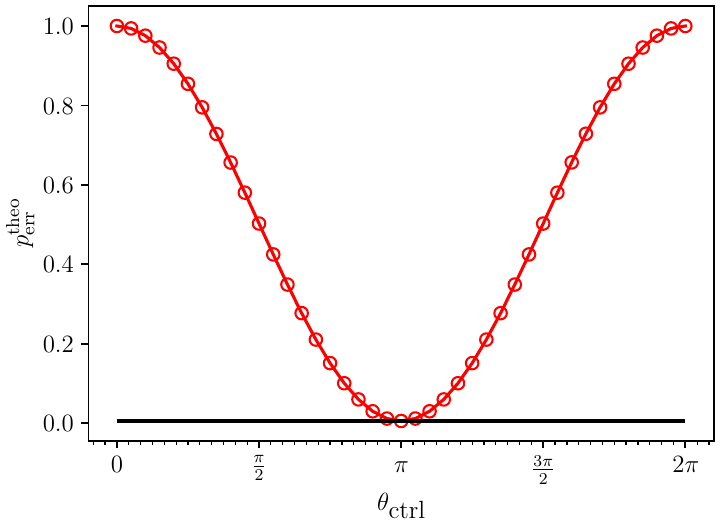}
    \caption{{Illustration of theoretical case error probability with respect to $\theta_\textrm{ctrl}$. The black line is referred to the error probability proposed in Eq.~\ref{eq:limiting-case-equation}.}}
    \label{fig:theor_prob_with_theta_ctrl}
\end{figure}

\begin{figure}[ht!]
	\centering
	\includegraphics[width = 0.8\linewidth]{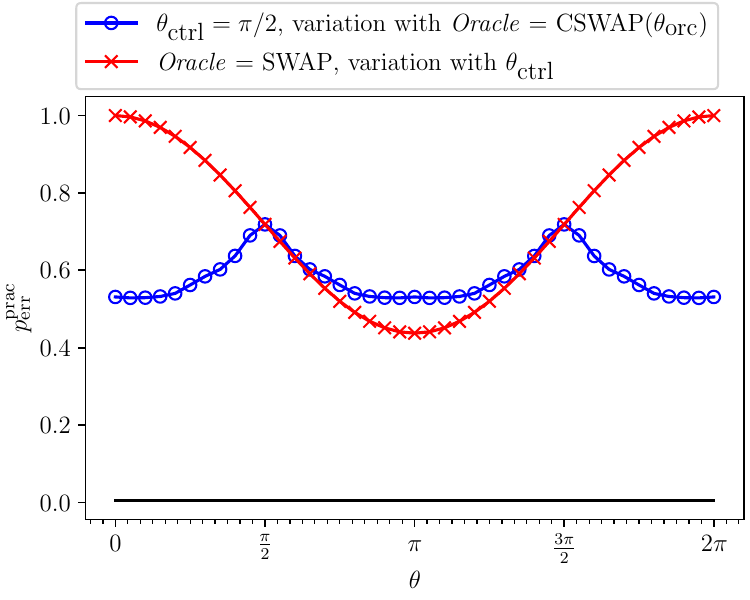}
    \caption{Illustration of variation of practical case error probability with respect to $\theta_\textrm{ctrl}$, when $\theta_\textrm{orc} = 0.0$ (blue line)  and $\theta_\textrm{orc}$, when $\theta_\textrm{ctrl} = \pi/2$ (red line). The black line refers to the error probability proposed in Eq.~\ref{eq:limiting-case-equation}.}
	\label{fig:prac_prob_with_thetas}
\end{figure}

For example, the unitary oracle as given in Eq.~\ref{eq:oracles_2} for the \texttt{Hypothesis 1} at $\theta_\text{ctrl}=0$ becomes $U_\text{orc}^{H_{1}}(0)=\mathbb{I}=U_\text{orc}^{H_{0}}(0)$, in that case, the two hypothesis becomes identical and the error probability $p_\textrm{err} =1$ which can be clearly seen in Fig.~\ref{fig:theor_prob_with_theta_ctrl} and~\ref{fig:prac_prob_with_thetas}. 
And as the $\texttt{Hypothesis 1}$ deviates from the $\texttt{Hypothesis 0}$, which can be achieved by fine-tuning $\theta_\text{ctrl}$, we see the error probability gradually decreases from $1$. 
At $\theta_\text{ctrl}=\pi$ when  $\texttt{Hypothesis 1}=\texttt{SWAP}$ the error probability obtains the limit reached by Eq.~\ref{eq:limiting-case-equation} (represented by the black horizontal line in Fig.~\ref{fig:theor_prob_with_theta_ctrl}). A brief discussion of Fig.~\ref{fig:theor_prob_with_theta_ctrl} and ~\ref{fig:prac_prob_with_thetas} is provided in Section~\ref{sec:numeric-results}.

So there is a clear indication that the distance between the oracles plays a crucial role in defining the error probability. Hence, to incorporate this dependence on the relative distinguishability of the hypotheses, we introduce a correction factor proportional to the process distance ($\Delta$) between the two oracles. This arises a modified version of Eq.\eqref{eq:limiting-case-equation} that takes the form:

%\begin{widetext}
\begin{align}
p_\textrm{err}^\textrm{prac} &= 1- \Delta\left[U_\textrm{orc}^{H_0}, U_\textrm{orc}^{H_1}\right](1-p_\textrm{err})\nonumber\\
&\xrightarrow[]{r>>1}1-\Delta\left[U_\textrm{orc}^{H_0}, U_\textrm{orc}^{H_1}\right]\left(1-\frac{1}{4rd^N}\right)
\label{eq:practical-case-equation}
\end{align}
%\end{widetext}

There are many choices for the distance $\Delta$ function between the hypotheses and need to be chosen based on the experimental and theoretical specifications of the application such as
\begin{enumerate}[nolistsep,noitemsep]
	\item \textit{Trace distance} which is defined by 
	\begin{equation}
	\Delta = \frac{1}{2}\textrm{Tr}|\rho_\textrm{orc}-\rho_\textrm{orc}^\textrm{alter}|,
	\end{equation}
	where $\rho$ represents the Choi representation of the unitary.
	\item \textit{Bures distance} which defined as 
	\begin{equation}
	\Delta = 2\left(1-\sqrt{F(\rho_\textrm{orc},\rho_\textrm{orc}^\textrm{alter})}\right),
	\end{equation}
	where $F$ quantifies the process fidelity between the Choi representation of oracles.
	\item \textit{Hilbert-Schmidt distance} which is defined by
\begin{equation}
	\Delta = \textrm{Tr}\left[\left(\rho_\textrm{orc} - \rho_\textrm{orc}^\textrm{alter}\right)^2\right].
\end{equation}
\end{enumerate}
But the numerical results show that the error probability after simulating the circuit in Fig.~\ref{fig:prac_prob_with_thetas} i.e. practical case error probability, $P_\text{err}^\text{prac}$, coincides with \textit{Hilbert-Schmidt distance}.

\begin{figure}[H]
	\centering
	\includegraphics[width = 0.8\linewidth]{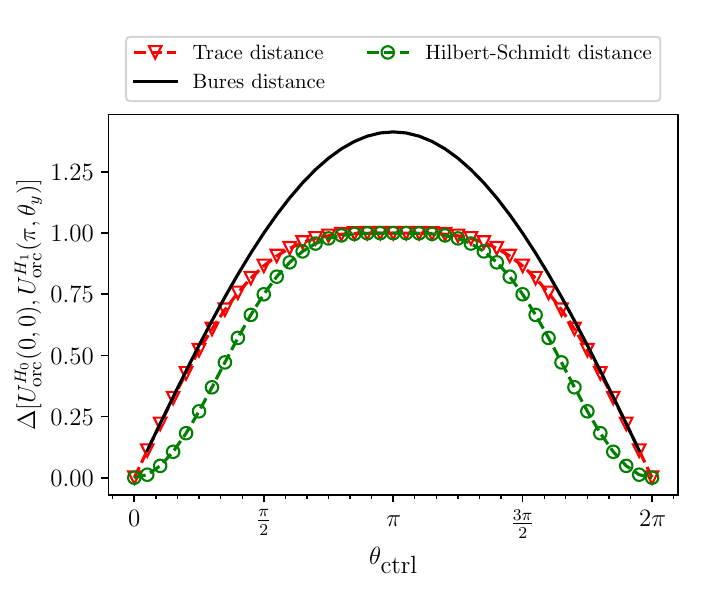}
	\caption{Illustration variation of different distance measures with respect to the oracle angle $\theta_\text{ctrl}$.}
	\label{fig:diff_dist_meas}
\end{figure}

\section{Numerical results}\label{sec:numeric-results}

To obtain numerical results, we consider 
IBM's open-source quantum computer simulator \texttt{Qiskit} was used to simulate the above-mentioned implementation.
At first, we choose the oracle unitary as given in Eq.~\ref{eq:oracles_2}
{where $q_\textrm{anc}$ is an ancilla qubit that allows us to control the strength of the \texttt{Hypothesis 1}. And the strength is dependent on parameter $\theta_\textrm{ctrl}$.
To obtain the specific hypothesis cases of $\mathbb{I}$ and \texttt{SWAP}, the above parameters are set to $U_\textrm{orc}^{H_{0}}(0)$ and $U_\textrm{orc}^{H_{1}}(\pi)$ respectively.
The dependence of the error probability of distinguishing two hypotheses with respect to the relative difference is enquired by varying the alternate hypothesis to
\begin{equation}
U_\textrm{orc}^{H_{0/1}} = 
\begin{cases}
U_\textrm{orc}^{H_{0}}(0), \\\\
U_\textrm{orc}^{H_{1}}(\theta_\textrm{ctrl}).
\end{cases}
\label{eq:alternate-hypothesis-theta-ctrl}
\end{equation}
}

{For the sake of better understanding, we compare the theoretical and practical case error probability, which is introduced in Eq.~\ref{eq:practical-case-equation}. For the theoretical scenario, we numerically calculate the Eq.~\ref{eq:practical-case-equation} by utilizing the \textit{Hilbert-Schmidt distance} between two hypotheses with respect to $\theta_\textrm{ctrl}$. Meanwhile, for the practical scenario utilize the introduced causal hypothesis testing circuit in Fig.~\ref{fig:full-perm-circuit} and get the error probability directly from the measurement outcomes.

{In Fig.~\ref{fig:theor_prob_with_theta_ctrl} we illustrate the theoretical scenario to evaluate the error probability.} {The black horizontal line refers to the results corresponding to the limiting case error probability in ref.~\cite{chiribella2019quantum}. We find that when the $\theta_\textrm{ctrl} = 0$, the \texttt{RX} gate that is controlling the SWAP oracle is not activated. This makes the alternate hypothesis (\texttt{Hypothesis 1}) $\mathbb{I}$. As our default hypothesis is already $\mathbb{I}$, we can not distinguish between the two hypotheses by any method. This gives theoretical case error probability (i.e. $p_\textrm{err}^\textrm{theo}$) 1.0. In the same way when $\theta_\text{ctrl}=\pi$ the SWAP oracle is activated by the \texttt{RX} control and we find the minimal probability of distinguishing between SWAP and $\mathbb{I}$, overlapping with the results from ref.~\cite{chiribella2019quantum}, in Eq.~\ref{eq:limiting-case-equation}.}
It is logical to examine the variation of process distance with respect to the oracle parameter $\theta_\text{ctrl}$. In Fig.~[\ref{fig:diff_dist_meas}] we illustrate the distance between the null hypothesis $U_\textrm{orc}^{H_{0}}(0,0)$ and the alternate hypothesis $U_\textrm{orc}^{H_{1}}(\pi,\theta_\text{ctrl})$ for a class of distance measure.

For the sake of experiments, we restrict ourselves to the observation of \textit{Trace distance}, \textit{Bures distance}, and the \textit{Hilbert-Schmidt distance}. It can be seen that the characteristics of \textit{Trace} and \textit{Hilbert-Schmidt} are similar since both of them fundamentally depend on the difference in the Choi representation of the oracles, whereas \textit{Bures distance} depends on the process distance.

{In the next experiment, we modify the unitary presented in Eq.~\ref{eq:oracles_2} to test the dependence on the error probability with other intermediary hypotheses.
We generate these intermediary hypotheses using a set of \texttt{parameterized CSWAP} gates generated by the decomposition implemented as 3 \texttt{iSWAP} with 3 interleaved \texttt{SX} gates on alternating qubits.
The general form of the oracle is given in Eq.~\ref{eq:oracles}.

\begin{widetext}
\begin{equation}
U_\textrm{orc}^{H_{0/1}}(\theta_\text{ctrl},\theta_\text{orc}) = \big[\textrm{CSWAP}(\theta_\text{orc}, q_\textrm{anc},A_i,B_i)^{\otimes N}\big]
\big[\textrm{RX}(q_\textrm{anc},\theta_\textrm{ctrl})\otimes\mathbb{I}(A_i)^{\otimes N}\otimes \mathbb{I}(B_i)^{\otimes N}\big],
\label{eq:oracles}
\end{equation}
\end{widetext}

To generate the Fig.\ref{fig:prac_prob_with_thetas} we make use of the parameterized controlled \texttt{SWAP} as the generator alternative hypothesis. 
}

{For this experiment, we used the quantum circuit and performed a state vector simulation. In Fig.~\ref{fig:prac_prob_with_thetas}, the red line shows the variation of the practical error probability with the \texttt{SWAP} gate. The difference with Fig.~\ref{fig:theor_prob_with_theta_ctrl} is due to the practical case being implemented using the sum of probabilities of distinct states. Meanwhile, the blue line shows the variation with respect to the \texttt{parameterized CSWAP} for the ancilla-control set to $\pi/2$. As expected, for the oracle angle of $\pi/2$, the \texttt{CSWAP} is the same as a \texttt{SWAP}.}

We observe that the results for the theoretical case and the numerically simulated practical case error probability are distinct (see the red lines in Fig.~\ref{fig:theor_prob_with_theta_ctrl} and Fig.~\ref{fig:prac_prob_with_thetas}). 
This is because of two factors: (i) the $p_\textrm{err}$ was introduced purely theoretically where the authors evaluated it for a maximally mixed alternate hypothesis. In a practical case (numerical simulation of the quantum circuit) it is not feasible to construct a maximally mixed oracle. This leads us to introduce the realistic case error probability. 
% In the practical scenario (see oracle presented in \eqref{eq:oracles}) the alternate hypothesis is implementable in currently widely available hardware and through the simulation, we show that the limiting case error probability in the practical scenario requires a slight modification. 
This modification is reflected through the introduction of the distance between the oracles under consideration i.e. $\Delta\left[U_\textrm{orc}^\textrm{alter},U_\textrm{orc}\right]$. 
% Based on this in Eq.\eqref{eq:practical-case-equation} we introduce a modified version of limiting case error probability that can be evaluated realistically on quantum hardware.
(ii) the maximum distinguishing measurement as suggested in the original formulation, requires knowledge of the hypothesis and the oracles, which cannot be known apriori.
While the theoretical process distance can be experimentally achieved via quantum process tomography, in the second experiment, we show the practical case error probability for Z-axis measurements without pre-rotations.

{Thus, these experiments demonstrate that: (i) the quantum circuit we  proposed in Fig.~\ref{fig:full-perm-circuit} can indeed be utilized for causal hypothesis testing, and (ii) the success probability of distinguishing two hypotheses is dependent on the process distance between the two oracles.}

% The numerical illustration suggests

\section{Application framework} \label{sec:application framework}

With the quantum kernel presented above, in this section, we discuss an application framework in the context of two consequential applications.

\subsection{Bioinformatics}

Applications of causal inference are widespread in bioinformatics.
Specifically, inferring a causal network is practiced in medical diagnostics and genomics.

For example, causal discovery in Alzheimer’s pathophysiology is studied in ref.~\cite{shen2020challenges} with 9 variables (13 with longitudinal data).
Similarly, for detecting causal regulatory interactions between genes, tools like \texttt{Scribe-py}~\cite{qiu2020inferring} are currently used.
Exemplary use cases of (i) transcription expression dynamics hierarchy of C. elegans' early embryogenesis and (ii) core regulatory network responsible for myelopoiesis are used for this research, with the latter graph consisting of 10 nodes.

The current generation of quantum processors supports 100s of qubits and is expected to scale to 1000s in a few years.
However, the challenge is the limitation of the decoherence time and gate errors, which bounds the runtime of the algorithm that can be effectively executed.
For a causal graph in the order of 10 causes/effects, a causal specificity bits of $d=1$ and with $N=100$, the estimation presented in Eq.\eqref{eq:qubit-requirement} is $6262$ qubits.
Keeping in mind the potential of near-term devices, pragmatic industrial cases of causal tomography will remain outside the reach of quantum computing in the near term.

% how to make the oracle? grover search?

\subsection{Artificial General Intelligence}

In the long term, quantum accelerated causal inference will be beneficial for artificial general intelligence.
Quantum accelerated AGI is still in its infancy. In ref.~\cite{catt2020gentle}, the authors proposed the AIXI-q reinforcement learning agent empowered with quantum counting. Meanwhile, ref.~\cite{sarkar2021estimating} proposed an exhaustive enumeration of all causal oracles (or, alternatively, all bounded-size programs of a Turing machine).
These techniques can develop synergies with the quantum accelerated causal tomography circuit as developed in this article.

In theoretical physics, automated science tools, specifically in the context of causal set theory, will also find the causal tomography framework of crucial use. 

\section{Conclusion} \label{sec:conclusion}

In this article, we extend the previously introduced (see ref.~\cite{chiribella2019quantum}) causal hypothesis testing formulation for the practical scenario. This led us to develop a scalable quantum gate-based algorithm which can be implemented in the available near-term quantum devices. Through the algorithm formulation, we empirically show that the limiting case error probability that is represented in ref.~\cite{chiribella2019quantum} requires modification. In our work, this modification is done by introducing process distance between the causal hypotheses, to the formulation of error probability. We term this modified version of error probability as \textit{practical case error probability} which is stricter than the limiting case. Additionally, our implementation enables an estimation of the pragmatic gate complexity of the causal tomography entangled pair indexing.

Furthermore, the proposed algorithm is implemented using the open-source quantum programming and simulation platform \texttt{qiskit}. As in ref.~\cite{chiribella2019quantum} it is shown that the quantum advantage holds for generalized probabilistic theory so the practical case that we present here is optimal in any scenario.

Our motivation for this project is driven by the increasing focus on causal inference in machine learning.
Besides monitoring the information flow in future quantum communication networks, as discussed in ref.\cite{chiribella2019quantum}, causal tomography is crucial for understanding the bounds of general intelligence and for bioinformatics use cases.
In our future work, we aim to apply our developed quantum accelerated causal tomography framework for these applications. 
% {Further work, finding better measurement for process distance.}

\begin{acknowledgments}
This project was initiated under the QIntern 2021 project ``Reinforcement Learning Agent for Quantum Foundations".
Therefore we would like to thank the organizers of the program and QWorld Association. 
We would also like to thank QIndia for providing us with a collaboration platform for the ideas motivating this research.
AK was partially supported by the Polish National Science Center (NCN) under the grant agreement $2019/33/B/ST6/02011$.
\end{acknowledgments}

\appendix

% \section{Circuit Implementation}
% Here we provide an illustration of the circuit that has been implemented to obtain the numerical results. Th dimension of the subsystem is kept at $4$ due to the hardware limitations.

% \section{Poster}
% Here attach the poster which has been presented in the \texttt{Causalworlds: The interface between quantum and relativistic causality, foundations and practicalities}. The link to the conference  \href{https://causalworlds.ethz.ch/}{https://causalworlds.ethz.ch/}.
%\begin{figure*}
%\centering
%\includegraphics[width = \linewidth]{poster/practical_error_probability.pdf}
  %  \caption{The poster.}
  %  \label{fig:my_label}
%\end{figure*}

\nocite{*}
\bibliography{aipsamp}% Produces the bibliography via BibTeX.

\end{document}